\documentstyle[12pt]{article}
\def\beq{\begin{equation}}
\def\eeq{\end{equation}}

\begin{document}

\begin{titlepage}
\begin{flushright}
BA-99-39 \\
May 1, 1999 \\
\end{flushright}

\begin{center}
{\Large\bf Proton Decay, Neutrino Oscillations And \\
Other Consequences From Supersymmetric SU(6) \\
With Pseudo-Goldstone Higgs
\footnote{Supported in part by  DOE under Grant No. DE-FG02-91ER40626
and by NATO, contract \\
~~~~~~~~~~~~number CRG-970149.}
}
\end{center}
\vspace{0.4cm}
\begin{center}
{\large Qaisar Shafi$^{a}$\footnote {E-mail address:
shafi@bartol.udel.edu} {}~and
{}~Zurab Tavartkiladze$^{b}$\footnote {E-mail address:
z\_tavart@osgf.ge} }
\vspace{0.5cm}

$^a${\em Bartol Research Institute, University of Delaware,
Newark, DE 19716, USA \\

$^b$ Institute of Physics, Georgian Academy of Sciences,
380077 Tbilisi, Georgia}\\
\end{center}

\vspace{0.5cm}

\begin{abstract}

We suggest a new  mechanism for naturally suppressing dimension five
baryon number violating in supersymmetric $SU(5+N)$
($N=0, 1,\cdots $) GUTs. The mechanism is realized through
suppression of $qqT$ type couplings, and is implemented by
introducing new `matter' multiplets belonging to
symmetric representations of $SU(5+N)$.
Together with the suppression of nucleon decay, these multiplets also
enable one to avoid the unwanted asymptotic mass relations
$m_s=m_{\mu }$, $\frac{m_d}{m_s}=\frac{m_e}{m_{\mu }}$.

As an example, we consider a $SU(6)$ model with
pseudo-Goldstone Higgs. By supplementing the model with an anomalous
${\cal U}(1)$ flavor symmetry, we also obtain a simple `all-order' solution
of the gauge
hierarchy problem and natural explanation of charged fermion mass
hierarchies and values of the CKM matrix elements. The proton life time
$\tau_p\sim 10^{2}\tau_p^{SU(5)}$~yr. is compatible with experiments, with
the
dominant decay being $p\to K\nu_{\mu, \tau}$. Thanks to the
$SU(6)$ symmetry, successful unification of the gauge couplings can be
retained, and the value of the strong coupling $\alpha_s(M_Z)$ can be
reduced to
$\simeq 0.12$.

Finally, we show how to accommodate the solar and atmospheric neutrino
data through the bi-maximal neutrino mixing scenario, with
maximal vacuum $\nu_e-\nu_{\mu, \tau }$  and large angle
$\nu_{\mu }-\nu_{\tau }$ oscillations.

\end{abstract}

\end{titlepage}

\section{Introduction}

One of the most outstanding problems of supersymmetric (SUSY) GUTs is the
question of
nucleon stability. Within the frameworks $SU(5)$ and $SO(10)$
GUTs the nucleon life time is about
$10^{29\pm 2}$~yr, to be
compared with the latest experimental bound
$\tau_N^{\rm exp}\stackrel {>}{_\sim }10^{32}$yr \cite{dat}.
This result is valid for the simplest versions of SUSY $SU(5)$,
$SO(10)$, etc., unless some mechanism for suppressing or eliminating the
dangerous baryon number violating $d=5$ operators is applied
\footnote{In SUSY theories such as flipped $SU(5)$, $SU(3)^3$,
flipped $SU(6)$ and $SU(4)_c\times SU(2)_L\times SU(2)_R$, the situation
is not so hard and it is easier to achieve the suppression of nucleon
decay by imposing some continuous or discrete ${\cal R}$-symmetries.
See \cite{flipsu5}-\cite{422} respectively.}.
In $SO(10)$, it seems, possible to resolve this problem,
through the missing VEV solution,with the VEV of the adjoint $45$-plet
Higgs in the $T_R$ direction \cite{babu, gia}.
Attempts to suppress dimension five of nucleon decay in the framework of
extended
(with $75$, $\overline{50}$, $50$ multiplets) SUSY $SU(5)$ theory were
suggested in \cite{psu50},
while in \cite{psu51}
the different mechanism,
tied to the pattern of fermion masses, was proposed.

In this paper we present a new mechanism for the suppression
of dimension five nucleon decay
in SUSY $SU(5+N)$ ($N=0, 1, \cdots $) GUTs. The mechanism is realized via
symmetry reasons, and a crucial role is played by symmetric pairs of
vector-like `matter' $\bar S+S$, which we introduce in the theory.
Through a special arrangement
of the couplings involving these `$S$'-plets, the $qqT$-type terms
can be suppressed to the needed level.  [Here $q$ and $T$ belong to
the (3,2) and (1,3) representations of
$SU(3)_c \times SU(2)_W$ respectively.]  Further, as pointed out in
\cite{psu5}, it turns out that these symmetric multiplets are important
not only for nucleon stability, but also for obtaining a realistic
pattern of fermion masses. Namely, thanks to these additional states,
the unwanted asymptotic relations $m_s=m_{\mu }$,
$\frac{m_d}{m_s}=\frac{m_e}{m_{\mu }}$ can be avoided, and a
reduced value, compared to SU(5) for $\alpha_s(M_Z)$ can be achieved.

For a transparent demonstration of the proposed mechanism we consider a SUSY
$SU(6)$ model, in which the doublet-triplet (DT) hierarchy is
achieved through the pseudo-Goldstone boson (PGB) mechanism. An anomalous
${\cal U}(1)$ symmetry, which we use, leads to an `all-order'
DT hierarchy \cite{golds1} and a desirable pattern of symmetry breaking.
We
consider
${\cal U}(1)$ as a flavor symmetry acting on the fermion generations and
gain a natural explanation of the hierarchies of charged fermion masses and
CKM matrix elements.

The scenario suggests a desirable value (for the SU(6) PGB model) of the
MSSM
parameter $\tan \beta$($\sim$ unity), retains $b-\tau $ unification and
avoids the analogous unwanted asymptotic relations for light generations.
Although the PGB $SU(6)$ scenarios considered earlier 
\cite{golds, golds1}, lead to other realistic implications, the problem of
nucleon stability had not been previously addressed within its framework
and was
an open question. Applying the proposed mechanism for suppression of
nucleon decay, we will see that the $SU(6)$ scenario which we consider
does not suffer from this problem.
The dominant nucleon decay mode is $p\to K\nu_{\mu, \tau  }$, and the life
time
is $\sim 10^2$ times larger then the minimal SUSY $SU(5)$ value.
Due to the peculiarity of $SU(6)$ gauge group and some additional
multiplets, the successful unification of three gauge couplings can
be retained and a reduced value for $\alpha_s(M_Z)$($\simeq 0.12$)
can be achieved.

Finally, for accommodating the solar and atmospheric neutrino data, we
consider the `bi-maximal' neutrino mixing scenario, in which the solar
neutrino puzzle is resolved through maximal $\nu_e-\nu_{\mu, \tau }$
vacuum oscillations, while the atmospheric neutrino deficit is due to large
$\nu_{\mu }-\nu_{\tau }$ mixing. Alternative scenarios for neutrino
oscillations are possible, as was
recently discussed in \cite{bimax}.

\section{Nucleon Decay and Mechanism for Suppression \\
 of $qqT$ Type Couplings in $SU(5+N)$ GUTs }

As pointed out some time ago in \cite{d5}, in supersymmetric GUTs (whose
minimal version is the $SU(5)$ gauge group), there is a new source for
baryon number violation. Namely, dimension five $(d=5)$ operators

\beq
{\cal O}_L=\frac{1}{M_T}qqql~,~~~~~
{\cal O}_R=\frac{1}{M_T}u^cu^cd^ce^c~,
\label{d5r}
\eeq
which emerge through the couplings

\beq
qqT+ql\bar T+u^ce^cT+u^cd^c\bar T+M_T\bar TT~,
\label{couplings}
\eeq
after integrating out the heavy color triplets $\bar T, T$.
For $M_T\sim 10^{16}$~GeV, the
dominant decay mode is $p\to K\nu_{\mu }$
\footnote{As it turns out, the dominant contribution to nucleon decay is
from the ${\cal O}_L$ operator, with the decay modes emerging through
${\cal O}_R$ being more suppressed.},
resulting in proton lifetime of
$10^{29\pm2}$
yr, which is embarrassingly small in comparison with
the latest experimental limit
$\tau_p^{\rm exp}\stackrel {>}{_\sim }10^{32}$yr \cite{dat}.

In this section we propose a new mechanism for suppressing the $d=5$
operators in the framework of supersymmetric $SU(5+N)$
GUTs, where $N=0,1,\dots $. For its breaking down to the
$SU(3)_c\times SU(2)_W\times U(1)_Y$ , it is sufficient to
introduce one adjoint ($\Sigma $), and $N$ pairs
of fundamental-antifundamental
($N\cdot(\bar H+H)\equiv\bar H^{(a)}+H^{(a)}$, $a=1,\dots , N$)
scalar
supermultiplets.

As far as the fermion sector is concerned, the
anomaly free chiral supermultiplets (which unify
quark-lepton superfields) of $SU(5+N)$ can be taken as follows
\footnote{For demonstration of the mechanism we will consider the simplest
choice of anomaly free
chiral `matter'. However, other choices are also possible.}:

\beq
A_{ij}+(N+1)\cdot \bar F^i~,
\label{matter}
\eeq
(for simplicity we consider only one generation)
where $i,j(=1,\dots ,N+5)$ are gauge indices, and
$A$ and $\bar F$ are the antisymmetric and (anti-)fundamental
representations respectively. In terms of $SU(5)$ these
multiplets contain, in addition to
$10+\bar 5$ superfields, $N$ pairs of vector-like $\bar 5+5$ and
$N(3N+1)/2$ singlet states. After the GUT symmetry breaks to $SU(3)_c\times
SU(2)_W\times U(1)_Y$
all the singlet and vector-like `matter' decouples and we are
left with the minimal chiral superfields in $10+\bar 5$:

\beq
10=(q,~u^c,~e^c)~,~~~~~\bar 5=(d^c~,l)~.
\label{chir}
\eeq
Since $A\supset 10\supset q$, and we are interested in $qqT$-type
couplings, we will consider the relevant couplings involving the
superfield $A$.

Let us introduce
a pair of symmetric  supermultiplets $S_{ij}+\bar S^{ij}$.
Together with other fragments, they  contain
$q+\bar q$ states respectively.
From the superpotential couplings

\beq
W(A, S)=A\Sigma \bar S+M_S\bar SS~,
\label{as}
\eeq
and assuming that 

\beq
\langle \Sigma \rangle \equiv M_G\gg M_S~,
\label{stabcond}
\eeq
one
can easily verify that $q_A$ decouples by forming a massive state with
$\bar q_S$, and the `light' state $q$ mainly resides in $S$:

\beq
S\supset q~,~~~~~~~A \stackrel{\supset }{_\sim} \frac{M_S}{M_G}q~,
\label{weights}
\eeq
The states $u^c$ and $e^c$ reside purely in $A$,

\beq
A\supset (u^c,~e^c)~.
\label{weight1}
\eeq

The operator which generates mass for up-type quark has the form
\beq
\frac{1}{M^{N+1}}(\Sigma S)_{mn}A_{pq}H^{(1)}_{i_1}\dots
H^{(N)}_{i_N}\frac{(\Sigma H^{(a)})_i}{M}\epsilon^{mnpqi_1\dots i_Ni}~,
\label{upqq}
\eeq
where convolution of $SU(5+N)$ group indices are indicated, and $M$ is
some cut-off mass scale. For the Yukawa coupling of up-type
quark we have to substitute the VEVs of appropriate scalar fields and also
extract from them the Higgs doublet $h_u$ (in general the $h_u$ state can
reside, with suitable weights in $H^{(a)}$ and also in $\Sigma$,
depending on the specifics of the model). Due to
(\ref{weights}), (\ref{weight1}) the states $q$, $u^c$ should be extracted
from $S$, $A$ respectively, and the corresponding Yukawa coupling will have
the form:

\beq
\lambda^Uqu^ch_u~.
\label{up0}
\eeq

For the $qqT$-type coupling, the light $q$ state should be extracted
from $A$, and since the latter contains it with a suppressed weight
(see (\ref{weights})), the $qqT$ coupling will also be suppressed:

\beq
\frac{M_S}{M_G}\lambda^UqqT~,
\label{qq}
\eeq
Consequently, the nucleon decay width induced from this coupling
(if $ql\bar T$ coupling has the `standard' form), will
be suppressed by a factor $\left(\frac{M_S}{M_G}\right)^2$ in comparison
to minimal $SU(5)$.
To obtain the
desired suppression it is enough to have $\frac{M_S}{M_G}\sim 10^{-2}$.
Therefore, with the help of $\bar S+S$ states we can get a natural
suppression of nucleon decay.
A crucial assumption for realizing sufficient nucleon stability is
(\ref{stabcond}), which
should be satisfied for the  mechanism described here to work.

It is worth noting that in addition to their role in the suppression of
nucleon decay,
the $\bar S+S$-plets are crucial also for avoiding the unwanted
asymptotic relations $\lambda_d^{\alpha }=\lambda_e^{\alpha }$ for the
`light'
generations \cite{psu5}. This relation is automatically violated
if the condition $\langle \Sigma \rangle \gg M_S$ is satisfied and the light
$q$, $e^c$ states reside in different multiplets (in the above example
in $S$ and $A$ respectively). If we wish, however, to retain the asymptotic
relation
$\lambda_b=\lambda_{\tau }$, the above mechanism should be applied
only to the first two generations. This means that the coupling $q_3q_3T$
is not suppressed, so that nucleon decay can also emerge from this term
through mixing with the `light' generations. Since the
mixings are small nucleon decay will still be
suppressed.  Whether the main contribution to the nucleon decay width
arises
from this effect, or from the direct couplings (\ref{qq}) depends
on the details of the model.

In concluding this section, let us note that since the mass scale $M_S$
of $\bar S+S$ states is below $M_G$, additional fragments
appearing in the $M_S-M_G$ region can change the running of the gauge
couplings, and
so care is needed for retaining the unification of the gauge couplings. In
the next
section we will consider the PGB $SU(6)$ scenario, which neatly meets this
constraint while leading to more stable nucleon than in $SU(5)$. As it turns
out, the dominant decay $p\to K\nu_{\mu, \tau }$
occurs through mixing of the `light' families with the third generation.

%
%
\section{Pseudo-Goldstone SU(6) Scenario
}

For a transparent demonstration of the mechanism described above,
we will consider the $SU(6)$ gauge theory which provides a
natural solution of the DT splitting problem through
the pseudo-Goldstone mechanism \cite{ino, golds, golds1, flipsu6}. It will
turn
out
that the peculiarities of some representations of $SU(6)$ allow the
possibility to retain the successful unification of the three
gauge couplings, even though the masses of additional symmetric
multiplets lie below the GUT scale.

\subsection{Higgs Sector: Symmetry Breaking and `All Order' DT Hierarchy}

In addition to $\Sigma(35)+\bar H(\bar 6)+H(6)$ Higgs supermultiplets
(which are necessary for $SU(6)$ breaking to
$SU(3)_c\times SU(2)_W\times U(1)_Y\equiv G_{321}$) we introduce
an additional singlet superfield $X$, and an anomalous ${\cal U}(1)$
symmetry.  The latter plays a crucial role in achieving an `all-order'
DT hierarchy \cite{golds1}, and helps provide a natural understanding of
the magnitudes of
charged
fermion masses and mixings. The ${\cal U}(1)$ charges of these
superfields are

$$
Q_{\Sigma }=0~,~~~~Q_{H}=r~,
$$
\beq
Q_{\bar H}=r'~,~~~~Q_X=1~,
\label{ch}
\eeq
where $r,r'$ must be taken in such a way that in the scalar
superpotential

\beq
W=\frac{\Lambda }{2}\Sigma^2+\frac{\lambda }{3}\Sigma^3+
M_P^3\left(\frac{X}{M_P}\right)^m\left(\frac{\bar HH}{M_P}\right)^n
\label{scsup},
\eeq

\beq
m+2n=25~
\label{mn},
\eeq
so that

\beq
r+r'=-\frac{m}{n}<0~.
\label{rr}
\eeq
Values for $m, n$ should be chosen in such a way that a lower order term in
(\ref{scsup}) is not allowed. One desirable choice is $(m, n)=(7, 9)$.

From the first two terms in (\ref{scsup}) there exists a non-vanishing SUSY
conserving solution for $\langle \Sigma \rangle $ along the direction

\beq
\langle \Sigma \rangle ={\rm Diag}
(1,~1,~1,~-2,~-2,~1)\cdot V
\label{dir}
\eeq
with

\beq
V=\frac{\Lambda }{\lambda } \equiv M_P\epsilon_G(\sim 10^{16}{\rm GeV})~.
\label{sol}
\eeq
From the last term in (\ref{scsup})
$\langle X\rangle =\langle H\rangle =\langle \bar H\rangle =0$
at this stage.

The situation will be changed if the ${\cal U}(1)$ symmetry happens to be an
anomalous
gauge symmetry, arising in effective field theories from strings.
The cancellation of anomalies occurs through the Green-Schwarz mechanism
\cite{gsh}. Due to the anomaly, the Fayet-Iliopoulos term

\beq
\xi \int d^4\theta V_A
\label{fi}
\eeq
is always generated \cite{fi} where, in string theory \cite{xi},

\beq
\xi =\frac{g_A^2M_P^2}{192\pi^2 }{\rm Tr}Q~.
\label{xi}
\eeq
The $D_A$ term will have the form

\begin{equation}
\frac{g_A^2}{8}D_A^2=\frac{g_A^2}{8}
\left(\Sigma Q_a|\varphi_a |^2+\xi \right)^2~,
\label{da}
\end{equation}
where $\varphi_a$ denotes a superfield which carries a nonzero `anomalous'
$Q_a$ charge. Assuming that $\xi <0$ (${\rm Tr}Q< 0$) and taking into
account (\ref{ch}), (\ref{rr}), the cancellation of (\ref{da}) fixes the
$\langle X\rangle $ VEV,

\beq
\langle X\rangle =\sqrt{-\xi }~.
\label{vevx}
\eeq
Note that for  ensuring a non-zero VEV for $X$, the important thing is not
the absolute sign of
$\xi $, but the (opposite) relative signs of $Q_X$ and $\xi $.
Further, we will take

\beq
\frac{\langle X\rangle }{M_P}=\frac{\sqrt{-\xi }}{M_P}\simeq 0.22
\simeq\left(\frac{m_{3/2}}{M_P}\right)^{1/23}~,
\label{vevx1}
\eeq
where $m_{3/2}$ is the gravitino mass($\sim 10^3$~GeV) and
$M_P=2.4\cdot 10^{18}$~GeV.

After SUSY breaking in minimal $N=1$ SUGRA theory, the soft bilinear
terms

\beq
V_{soft}^m=m_{3/2}^2\left(|H|^2+|\bar H|^2 \right)
\label{soft}
\eeq
emerge, and taking into account (\ref{scsup}), (\ref{vevx1}), one obtains
nonzero
VEVs $\langle H\rangle $, $\langle \bar H\rangle $ (along the $SU(5)$
singlet direction):

\beq
\frac{\langle H\rangle }{M_P}\sim \frac{\langle \bar H\rangle }{M_P}
\equiv \frac{v}{M_P}
\sim \frac{\langle X\rangle }{M_P}\equiv \epsilon \simeq
\left(\frac{m_{3/2}}{M_P}\right)^{1/23}\simeq 0.22~.
\label{vevh}
\eeq

It will turn out $\epsilon $ is an important expansion parameter for
understanding the hierarchies among the Yukawa couplings and the magnitude
of the CKM matrix elements. It is clear that in attempting to express
these hierarchies through  appropriate powers of $\epsilon $,
the expansion operator should be an $SU(6)$ singlet. Without the 
superfield
$X$ we would only have $\bar HH/M_P^2\sim \epsilon^2$,
which is too
small. Also, as we will see in sect. $(3.2)$, odd powers of 
$\epsilon $ are also needed. Therefore, the introduction of the singlet
superfield
$X$ is crucial in our model.

The superpotential in (\ref{scsup}) has
$SU(6)_{\Sigma }\times U(6)_{H+\bar H}$ global symmetry which, by
(\ref{dir}), (\ref{vevh}) is broken down to
$[SU(4)_c\times SU(2)_W\times U(1)']_{\Sigma }\times U(5)_{H+\bar H}$,
while the $SU(6)$ gauge symmetry in broken to
$SU(3)_c\times SU(2)_W\times U(1)_Y$. It is easy to verify that the
pseudo-Goldstone states, `massless' if SUSY is unbroken, are a
doublet-antidoublet pair which can be identified with the
MSSM Higgs doublets\footnote{For more detailed discussions see
\cite{ino, golds}.}. The states

\beq
h_u=\frac{vh_u^{(\Sigma )}-3Vh_u^{(H)}}{\sqrt{v^2+9V^2}}~,~~~
h_d=\frac{vh_d^{(\Sigma )}-3Vh_d^{(H)}}{\sqrt{v^2+9V^2}}~,
\label{pgd}
\eeq
are `massless' pseudo-Goldstones, while their orthogonal superpositions

\beq
h_u^G=\frac{vh_u^{(\Sigma )}+3Vh_u^{(H)}}{\sqrt{v^2+9V^2}}~,~~~
h_d^G=\frac{vh_d^{(\Sigma )}+3Vh_d^{(H)}}{\sqrt{v^2+9V^2}}~,
\label{golds}
\eeq
are genuine Goldstones, `eaten up' by the appropriate gauge fields.
From (\ref{pgd}), (\ref{golds}), taking into account (\ref{sol}) and
(\ref{vevh}),
one can easily verify that the physical doublets reside in $\Sigma $
and $H, \bar H$ as follows:

\beq
\Sigma \supset (h_u,~h_d)~,~~~~
H\supset 3\frac{\epsilon_G}{\epsilon }h_u~,~~~~
\bar H\supset 3\frac{\epsilon_G}{\epsilon }h_d~.
\label{weight2}
\eeq

Note that the $SU(6)_{\Sigma }\times U(6)_{H+\bar H}$ global symmetry
violating higher
order operator $\Sigma X^m\left(\bar HH\right)^n$ gives the desirable
contribution ($\sim100$~GeV) to the $\mu $-term.
In summary, the anomalous ${\cal U}(1)$ symmetry helps provide an
`all-order'
solution of the DT splitting problem
in the framework of PGB $SU(6)$ model.

\subsection{Charged Fermion Masses and Mixings}

The `matter' sector of $SU(6)$ contains the anomaly free `matter'
supermultiplets $(15+\bar 6+\bar 6')_{\alpha }$ ($\alpha =1, 2, 3$
is a family index). For generating the top quark mass at
renormalizable level, we also introduce a $20$-plet of $SU(6)$. In order for
the
proton decay suppression mechanism (discussed in section 2) to
work, we introduce two pairs of symmetric supermultiplets
$(\overline 21+21)_{1,2}$. These fields will be crucial also for getting
a realistic pattern of charged fermion masses. In terms of $SU(5)$, the
`matter' superfields decompose as:

$$
\bar 6=\bar 5+1~,~~~~~15=10+5~,
$$
\beq
20=10+\overline{10}~,~~~21=15+5+1~.
\label{dec}
\eeq
The decomposition of $\overline{21}$ is analogous to $21$.

Here we will use the anomalous ${\cal U}(1)$  as a flavor symmetry in order
to
gain an understanding of hierarchies of Yukawa couplings and CKM matrix
elements. [Let us note that anomalous Abelian factors are
widely used in the literature for this purpose \cite{u1}.] The
transformation properties of
`matter' superfields under ${\cal U}(1)$ are presented
in Table (\ref{t:fer}). The relevant couplings
are:

\begin{table}
\caption{Transformation properties of `matter' superfields
under ${\cal U}(1)$ symmetry}
\label{t:fer}
$$\begin{array}{|c|c|c|c|c|c|c|c|c|}
\hline
& & & & & & & & \\
& 20  & 15_i& \bar 6_1~,\bar 6_1'&
\bar 6_{2,3}~,\bar 6_{2,3}'&
21_1  & \overline{21}_1 &
21_2  &\overline{21}_2
 \\
& & & & & & & &\\
\hline
& & & & & & & & \\
{\cal U}(1)&0  &i-3-r &-3-2r' &
-1-2r' &-2-r  &2+r  &
-1-r &1+r  \\
& & & & & & & & \\
\hline
\end{array}$$
\end{table}

$$
20\Sigma 20 +20\left(15_3+\frac{X}{M_P}15_2+\frac{X^2}{M_P^2}15_1
\right)H+
$$
\beq
+ 20\left(\bar 6_3+\bar 6_2+\frac{X^2}{M_P^2}\bar 6_1+
(\bar 6_i\to \bar 6_i')
 \right)\Sigma \frac{X\bar H^2}{M_P^3}
\label{20}
\eeq

\begin{equation}
\begin{array}{ccc}
 & {\begin{array}{ccc}
\hspace{-5mm}\bar 6_1(\bar 6_1')& \,\,~\bar 6_2(\bar 6_2')
& \,\,~\bar 6_3(\bar 6_3')~~~~~~~~~~~~~~~~~

\end{array}}\\ \vspace{2mm}
\begin{array}{c}
15_1 \\ 15_2 \\ 15_3
 \end{array}\!\!\!\!\! &{\left(\begin{array}{ccc}
\,\,\left(\frac{X}{M_P}\right)^5
&\,\,\left(\frac{X}{M_P}\right)^3 &
\,\,\left(\frac{X}{M_P}\right)^3
\\
\,\,\left(\frac{X}{M_P}\right)^4 &
\,\,\left(\frac{X}{M_P}\right)^2 &
\,\,\left(\frac{X}{M_P}\right)^2
 \\
\,\,\left(\frac{X}{M_P}\right)^3 &
\,\,\frac{X}{M_P} &\,\,\frac{X}{M_P}
\end{array}\right)\frac{\bar HH}{M_P^2}(1+\frac{\Sigma }{M_P})
\bar H }~,
\end{array}  \!\!
\label{15-6}
\end{equation}

\begin{equation}
\begin{array}{cc}
 & {\begin{array}{ccc}
\bar 6_1(\bar 6_1')&\,\,~\bar 6_2(\bar 6_2')&
\,\,~\bar 6_3(\bar 6_3')~~~~~~~~~~~~~~~~~~
\end{array}}\\ \vspace{2mm}
\begin{array}{c}
21_1\\ 21_2

\end{array}\!\!\!\!\! &{\left(\begin{array}{ccc}
\,\, \left(\frac{X}{M_P}\right)^5 &
\,\,  \left(\frac{X}{M_P}\right)^3 &
\,\, \left(\frac{X}{M_P}\right)^3
\\
\,\, \left(\frac{X}{M_P}\right)^4 &
\,\,\left(\frac{X}{M_P}\right)^2 &
\,\, \left(\frac{X}{M_P}\right)^2
\end{array}\right)\frac{\bar HH}{M_P^2}(1+\frac{\Sigma }{M_P})
\bar H }~,
\end{array}  \!\!
\label{21-6}
\eeq

\begin{equation}
\begin{array}{cc}
 & {\begin{array}{ccc}
15_1&\,\,15_2&\,\,15_3~~~
\end{array}}\\ \vspace{2mm}
\begin{array}{c}
\overline{21}_1\\ \overline{21}_2

\end{array}\!\!\!\!\! &{\left(\begin{array}{ccc}
\,\, 1~~&
\,\,  0~~ &\,\, 0
\\
\,\, \frac{X}{M_P} ~~ &\,\,1~~&\,\, 0~
\end{array}\right)\Sigma }~,
\end{array}  \!\!~
\begin{array}{cc}
 & {\begin{array}{cc}
21_1&\,\,
~21_2~~~~~
\end{array}}\\ \vspace{2mm}
\begin{array}{c}
\overline{21}_1 \\ \overline{21}_2

\end{array}\!\!\!\!\! &{\left(\begin{array}{ccc}
\,\, 1~~
 &\,\,0
\\
\,\, \frac{X}{M_P}~~
&\,\,1
\end{array}\right)M_S.
}
\end{array}
\label{21s}
\end{equation}

Without loss of generality one can choose a basis in which (\ref{20})
has the form

\beq
20\Sigma 20+2015_3H+\frac{X}{M_P^3}20\bar 6_3\Sigma \bar H^2~.
\label{201}
\eeq
Through this redefinition the hierarchical structures of
(\ref{15-6})-(\ref{21s}) will
not be changed. Assuming that $\langle \Sigma \rangle \gg M_S$ and taking
into account (\ref{21s}), (\ref{201}), we see that $10_{15_3}$ (after
forming
a massive state with $\overline{10}_{20}$) decouples. Also the states
$q_{15_{1,2}}$ $\bar q_{\overline{21}_{1,2}}$ decouple. The
appropriate `light' chiral fragments reside as follows:

$$
~15_1 \supset q_1,~~~15_2 \supset q_2~,~~~20 \supset (q, u^c, e^c)_3
$$,
$$
15_1\stackrel {\supset}{_\sim }\frac{M_S^{(1)}}{M_G}q_1~,~~~~~~
15_2 \stackrel {\supset}{_\sim} \frac{M_S^{(2)}}{M_G}\epsilon q_2~,
$$
\beq
15_1 \supset (u^c, e^c)_1~,~~~~~15_2 \supset (u^c, e^c)_2~,
\label{weight3}
\eeq
where $M_S^{(1,2)}$ are the eigenvalues of the second matrix in (\ref{21s}).

Let us begin with the masses of charged leptons and down quarks.
From (\ref{15-6}), three
$\bar 5'_{\alpha }$ from $\bar 6'_{\alpha }$ couple with
$5_{\alpha }$ states from $15_{\alpha }$.
To get realistic pattern of fermion masses, the light $(d^c, l)_{\alpha }$
states should reside in $\bar 6_{\alpha }$. On the other hand,
from (\ref{21-6}) two states of $(d^c, l)_{\bar 6}$ can form massive states
with
$(\bar d^c, \bar l)_{21_{1,2}}$,
and decouple. This would mean that the light $(d^c, l)_{1,2}$ states reside
in
$\overline{21}_{1,2}$. To avoid this, we introduce two pairs of superfields
$[\overline{\chi }(\bar 6)+\chi (6)]_{1,2}$ with ${\cal U}(1)$ charges:

$$
Q_{\overline{\chi }_1}=2+r-r'~,~~~~~Q_{\chi_1}=-2-2r~,
$$
\beq
Q_{\overline{\chi }_2}=1+r-r'~,~~~~~Q_{\chi_2}=-1-2r~.
\label{charges}
\eeq
Then, from the couplings
\beq
W_{21,\chi}=21_i\overline{\chi }_i\bar H+
\overline{21}_i\chi_i H~,
\label{21-xi}
\eeq
the states $(\bar 5+5)_{1,2}$ (from $(\overline{21}+21)_{1,2}$)
became super heavy by coupling with
$(5_{\chi }+\bar 5_{\overline{\chi }})_{1,2}$ respectively.
Therefore, the $21$-plets affect only the appropriate $q$ states, and we
have\footnote{Without loss of generality we can assume that the light
$(d^c, l)_{\alpha }$ states reside in $\bar 6_{\alpha }$, while the
fragments from $\bar 6'_{\alpha }$ are superheavy.}
\beq
\bar 6_{\alpha }\supset (d^c, l)_{\alpha }~.
\label{weight4}
\eeq

The couplings (\ref{15-6}), (\ref{21-6}) are responsible for the generation
of
charged
lepton and down quark masses of the light generations respectively.
If $\Sigma $ in (\ref{15-6}) is not used, the charged leptons (of
light
generations) will remain massless since the $e^c$ states will have couplings
with superpositions (carrying quantum numbers of $l$) which are already
decoupled together with $\bar l_{15}$ states. Thus, $\Sigma $ in
(\ref{15-6})
is crucial, and $h_d$ should be extracted from it. Taking into account all
this, from (\ref{15-6}), (\ref{21-6}) and the last term in (\ref{201}), for
the mass
matrices of charged leptons and down quarks  we obtain:

\begin{equation}
\begin{array}{ccc}
 & {\begin{array}{ccc}
\hspace{-5mm}~l_1~~& \,\,l_2~~ & \,\,l_3~~

\end{array}}\\ \vspace{2mm}
\begin{array}{c}
e^c_1 \\ e^c_2 \\ e^c_3
 \end{array}\!\!\!\!\! &{\left(\begin{array}{ccc}
\,\,\epsilon^5 ~~ &\,\,\epsilon^3~~ &
\,\,\epsilon^3
\\
\,\,\epsilon^4 ~~  &\,\,\epsilon^2~~ &
\,\,\epsilon^2
 \\
\,\,0~~ &\,\,0~~ &\,\,1
\end{array}\right)\epsilon^3h_d }~,
\end{array}  \!\!  ~~~~~
\label{down}
\end{equation}

\begin{equation}
\begin{array}{ccc}
 & {\begin{array}{ccc}
\hspace{-5mm}~d^c_1~& \,\,d^c_2~& \,\,d^c_3~

\end{array}}\\ \vspace{2mm}
\begin{array}{c}
q_1 \\ q_2 \\ q_3
 \end{array}\!\!\!\!\! &{\left(\begin{array}{ccc}
\,\,\epsilon^5 ~~ &\,\,\epsilon^3~~ &
\,\,\epsilon^3
\\
\,\,\epsilon^4 ~~  &\,\,\epsilon^2~~ &
\,\,\epsilon^2
 \\
\,\,0~~ &\,\,0~~ &\,\,1
\end{array}\right)\epsilon^3h_d }~.
\end{array}  \!\!  ~~~~~
\label{down}
\end{equation}

Upon diagonalization, we find

\beq
\lambda_{\tau }\sim \epsilon^3~,~~~
\lambda_e:\lambda_{\mu }:\lambda_{\tau }\sim
\epsilon^5:\epsilon^2:1~,
\label{mlept}
\eeq

\beq
\lambda_b\sim \epsilon^3~,~~~
\lambda_d:\lambda_s:\lambda_b\sim
\epsilon^5:\epsilon^2:1~.
\label{mdown}
\eeq

Taking into account that the $\tau $ lepton and $b$ quark Yukawas are
generated from the last term of (\ref{201}) we will have

\beq
\lambda_b=\lambda_{\tau }\sim \epsilon^3\simeq 10^{-2}~,~~~~~
\label{btau}
\eeq
which means that the MSSM parameter $\tan \beta $ is of order unity. This
regime is also preferable for the PGB scenario \cite{tan}.
As far as the  light generations are concerned, due to fact that $e^c$
and $q$ states come from different ($15$ and $21$ respectively)
multiplets, the unwanted asymptotic relations $\lambda_e=\lambda_d$,
$\lambda_{\mu }=\lambda_s$ are avoided.

Turning to the up quark sector, from the first term in (\ref{201}) and
taking into account (\ref{weight3}), we have $\lambda_t\sim
1$.
The $u$ and $c$ quark masses are generated through the operators:

\beq
\frac{\Sigma^2 }{M_1}21_115_1\left(\frac{X}{M_P}\right)^4\frac{H^2}{M_P}+
\frac{\Sigma^2 }{M_2}21_215_2\left(\frac{X}{M_P}\right)^2\frac{H^2}{M_P}~,
\label{op}
\eeq
from which, taking into account (\ref{weight3}) and assuming $M_1\simeq
M_G$,
$M_2\simeq M_G\epsilon $, one obtains $\lambda_u\sim \epsilon^6$,
$\lambda_c\sim \epsilon^3$. The operators in (\ref{op}) can be obtained
through the exchange of pairs of
$[\overline{\Psi }(\overline{15})+\Psi(15)]_{1,2}$, with masses
$M_1$, $M_2$ respectively, and with ${\cal U}(1)$ charges

\beq
Q_{\Psi_1 }=-Q_{\overline{\Psi }_1}=2-r~,~~~~
Q_{\Psi_2 }=-Q_{\overline{\Psi }_2}=1-r~.~~~~
\label{chargepsi}
\eeq
The relevant couplings are

\begin{equation}
\begin{array}{cc}
 & {\begin{array}{ccc}
\overline{\Psi }_1~~~~~~&\,\,\overline{\Psi }_2~~~~~~~~~
\end{array}}\\ \vspace{2mm}
\begin{array}{c}
21_1\\ 21_2

\end{array}\!\!\!\!\! &{\left(\begin{array}{cc}
\,\, \left(\frac{X}{M_P}\right)^2~~&
\,\,  \frac{X}{M_P}
\\
\,\, \frac{X}{M_P} ~~ &\,\,1~
\end{array}\right)\left(\frac{X}{M_P}\right)^2\Sigma }~,
\end{array}  \!\!~
\begin{array}{cc}
 & {\begin{array}{cc}
\Psi_1&\,\,
~\Psi_2~~~~~~~~
\end{array}}\\ \vspace{2mm}
\begin{array}{c}
15_1 \\ 15_2

\end{array}\!\!\!\!\! &{\left(\begin{array}{ccc}
\,\, 1~~
 &\,\,\frac{X}{M_P}
\\
\,\, 0~~
&\,\,1
\end{array}\right)\frac{H^2}{M_P}\Sigma~.
}
\end{array}
\label{psis}
\end{equation}

\begin{equation}
\begin{array}{cc}
 & {\begin{array}{cc}
\hspace{-5mm}~\Psi_1~& \,\,~~~\Psi_2

\end{array}}\\ \vspace{2mm}
\begin{array}{c}
\overline{\Psi }_1 \\ \overline{\Psi }_2
 \end{array}\!\!\!\!\! &{\left(\begin{array}{ccc}
\,\,M_1 ~~ &\,\,M'\frac{X}{M_P}
\\
\,\,0~~  &\,\,M_2
\end{array}\right) }~,
\end{array}  \!\!  ~~~~~
\label{masspsi}
\end{equation}
For $M_1\sim M_G$, $M_2\sim \epsilon M_G $,
$M'\stackrel{_<}{_\sim}\epsilon M_G $ integration of
$\overline{\Psi}+\Psi $ states (and first term of (\ref{201}))
yields the following mass matrix

\begin{equation}
\begin{array}{ccc}
 & {\begin{array}{ccc}
\hspace{-5mm}~u^c_1~& \,\,u^c_2~ & \,\,u^c_3

\end{array}}\\ \vspace{2mm}
\begin{array}{c}
q_1 \\ q_2 \\ q_3
 \end{array}\!\!\!\!\! &{\left(\begin{array}{ccc}
\,\,\epsilon^6 ~~ &\,\,\epsilon^4~~ &
\,\,0
\\
\,\,0 ~~  &\,\,\epsilon^3~~ &
\,\,0
 \\
\,\,0~~ &\,\,0~~ &\,\,1
\end{array}\right)h_u }~,
\end{array}  \!\!  ~~~~~
\label{up}
\end{equation}
from which one can obtain

\beq
\lambda_t\sim 1~,~~~
\lambda_u:\lambda_c:\lambda_t\sim
\epsilon^6:\epsilon^3:1~.
\label{mup}
\eeq

From (\ref{down}), (\ref{up}) we have for the CKM matrix elements

\beq
V_{us}\sim \epsilon~,~~~~V_{cb}\sim \epsilon^2~,~~~~
V_{ub}\sim \epsilon^3~,
\label{ckm}
\eeq
which have the desired magnitudes!

In summary, by suitably implementing a flavor ${\cal U}(1)$
symmetry we have obtained a natural understanding of hierarchies of Yukawa
couplings and mixing angles of charged fermions in the framework of
PGB $SU(6)$ scenario.

\subsection{Unification and Value of $\alpha_s(M_Z)$}

In our scenario we have assumed that
$\langle \Sigma \rangle \gg M_S$ ($M_S$ is the mass scale of $21$-plets).
This assumption is important not only for obtaining the desired pattern of
fermion masses, it also is crucial for the suppression of nucleon decay. On
the other hand, it turns out that the states

\beq
2\cdot [(6,1)+(1,3)+(\bar 6,1)+(1,\bar 3)]
\label{states}
\eeq
(in terms of $SU(3)_c\times SU(2)_W$) will lie below the GUT scale, and if
$M_S$ is much lower than $M_G$, this will ruin the unification at $M_G$ of
the three
gauge couplings. However, the PGB $SU(6)$ scenario provides an
elegant possibility for retaining unification.
For this, we introduce two states $20_1$, $20_2$, which do not
couple with `matter' superfields\footnote{This is easily achieved if
for $20_{1,2}$, we either impose their `own' ${\cal Z}_2$ parity, or assume
that
they do not transform under ordinary `matter' parity.} and appear in the
terms:

\beq
W_{20}=\Sigma 20_i20_j+M_{20}20_120_2~,
\label{w20}
\eeq
(note that $20_120_1$ and $20_220_2$ couplings vanish due to $SU(6)$
symmetry).

Having $\langle \Sigma \rangle $ along the direction given in (\ref{dir}),
it is easy to
verify from the first coupling in (\ref{w20}) that the fragments
$(6,2)_{1,2}$
(in terms of $SU(4)_c\times SU(2)_W$) remain massless, while the remaining
components
get masses of order $M_G$. From the last coupling in (\ref{w20}) arises the
mass term $M_{20}(6,2)_1\cdot (6,2)_2$. In terms of
$SU(3)_c\times SU(2)_W$

\beq
(6,2)=(3,2)+(\bar 3,2)\equiv q'+\bar q'~.
\label{comp}
\eeq
Assuming that $M_{20}\simeq M_S$, it is easy to see that the states
$(q'+\bar q')_{1,2}$ from $20_{1,2}$, together with
fragments in (\ref{states}),
constitute two complete pairs of $SU(5)$ $15+\overline{15}$, and therefore
the picture of unification will not be altered in the one loop
approximation. As we will see in next section, for an acceptable suppression
of nucleon decay it is enough to have $M_S/M_G\sim 10^{-2}$.
A factor $10^{-2}$ for the suppression of $(6,2)_{1,2}$-plets masses
relative to $M_G$ is natural in the sense that
without the last term (\ref{w20}), they gain mass($\sim 10^{-2}M_G$)
through the coupling $\frac{\Sigma^2 }{M_P}20_120_2$.

It is worth pointing out that this scenario even allows the possibility of
obtaining a
reduced value (compared to $SU(5)$) for $\alpha_s(M_Z)$. Note that for
pairs in (\ref{states})
we have two mass eigenvalues $M_S^{(1,2)}$ (emerging from diagonalization
of the second matrix in (\ref{21s})). Assuming that $M_S^{(1)}\simeq
M_{20}$,
$M_S^{(2)}\simeq M_{20}/2$, one pair of $(6,1)+(\bar 6,1)$ and
$(1,3)+(1,\bar 3)$ will lie below the $M_{20}$ scale. For $\alpha_s(M_Z)$
this gives \cite{psu5}

\beq
\alpha_s^{-1}=\left(\alpha_s^{-1}\right)^0+
\frac{3}{2\pi }\ln \frac{M_{20}}{M_S^{(2)}}~,
\label{alps}
\eeq
where $\alpha_s^0$ is the value of the strong coupling at $M_Z$ in minimal
SUSY $SU(5)$
which is calculated in two loop approximation and includes
SUSY and heavy threshold corrections. Taking $\alpha_s^0=1/0.126$
\cite{lan} from (\ref{alps}),
we obtain $\alpha_s\simeq 0.121$, which is in good agreement with the world
average value \cite{dat}.

\subsection{Nucleon Decay in PGB SU(6)}

In this section we will investigate the all important issue of nucleon decay
in the PGB $SU(6)$
scenario. It will turn out that the proton life time is consistent with the
latest
Superkamiokande data, and the dominant decays occur through mixing of the
two
light families with the third generation. From
(\ref{21-6}), (\ref{201})
and also from (\ref{weight3}), we observe that the $q{\cal B}l\bar T$ type
couplings have the same hierarchical structure as the down quark mass
matrix in (\ref{down}). Taking into account (\ref{201}),
(\ref{weight3}) and the
assumption $M_S^{(1,2)}/M_G\equiv \eta \sim 10^{-2}$, the
integration of $(\overline{\Psi }+\Psi )_{1,2}$ states
(see (\ref{psis}), (\ref{masspsi})) yields the
$q{\cal A}qT$ type couplings.

\begin{equation}
\begin{array}{ccc}
 & {\begin{array}{ccc}
\hspace{-5mm}~q_1~~~& \,\,q_2 & \,\,~~~q_3

\end{array}}\\ \vspace{2mm}
\begin{array}{c}
q_1 \\ q_2 \\ q_3
 \end{array}\!\!\!\!\! &{\left(\begin{array}{ccc}
\,\,\eta \epsilon^6 ~~ &\,\,\eta \epsilon^4~~ &
\,\,0
\\
\,\,\eta \epsilon^4 ~~  &\,\,\eta \epsilon^3~~ &
\,\,0
 \\
\,\,0~~ &\,\,0~~ &\,\,1
\end{array}\right)T }~.
\end{array}  \!\!  ~~~~~
\label{qqT}
\end{equation}

Neglecting the third generation for the time being, we see that the nucleon
decay
amplitude is suppressed by $\eta \sim 10^{-2}$, which would lead
to $\tau_p\sim 10^4\cdot \tau_p^{SU(5)}$ ($\tau_p^{SU(5)}$ denotes the
proton lifetime in minimal SUSY $SU(5)$). Since the coupling
$q_3q_3T$ is not suppressed and there exist mixings between the appropriate
light and heavy states, we should investigate proton life time in more
detail.

The decay channel involving the neutrino occurs through operator

\begin{equation}
{\cal O}=x\cdot (u d_{\alpha })(d_{\gamma }\nu_{\beta })~,
\label{op1}
\end{equation}
which emerges from the exchange of color triplet higgsinos.  Here

$$
x=\alpha \left(
-(L_d^{\dag }{\cal B}L_e)_{\gamma \beta}
(L_u^{\dag }{\cal A} L_d^{*})_{\delta \rho }
\hat{V}_{\delta \alpha }\hat{V}^{\dag }_{\rho 1}
+
(L_u^{\dag }{\cal A} L_d^{*})_{1\alpha }(L_u^{\dag }{\cal B}
L_e)_{\delta \beta }\hat{V}_{\delta \gamma}-
\right.
$$
\begin{equation}
\left.
~~~~~~~-(L_d^{\dag }{\cal A} L_u^{*})_{\gamma \delta }
(L_d^{\dag }{\cal B}L_e)_{\rho \beta }
\hat{V}_{\delta \alpha } \hat{V}^{\dag }_{\rho 1}
-
(L_d^{\dag }{\cal A} L_u^{*})_{\alpha \delta }
(L_u^{\dag }{\cal B} L_e)_{1\beta }\hat{V}_{\delta \gamma }
\right)~,
\label{x0}
\end{equation}
$\alpha $ is a family independent factor, $\hat{V}$ is the CKM matrix

\beq
\hat{V}=L_u^TL_d^*~,
\label{ckmmat}
\eeq
and $L_e$, $L_d$, $L_u$
are unitary matrices which rotate the corresponding left handed states,
transforming  them from the flavor to the mass eigenstate basis.
Substituting (\ref{ckmmat}) in (\ref{x0}), one obtains

\beq
x=-2\alpha (L_d^{\dag }{\cal A} L_d^{*})_{\alpha \gamma }
(L_u^{\dag }{\cal B} L_e)_{1\beta }~.
\label{x1}
\eeq
Taking into account  ${\cal A}_{33}=\lambda_t$ and
$(L_d)_{13}=\hat{V}_{ub}^*$, $(L_d)_{23}=\hat{V}_{cb}^*$ (see forms of
(\ref{down}) and (\ref{up})), for $p\to K\nu_{\mu, \tau }$ decay width we
have

\beq
\Gamma (p\to K\nu_{\mu, \tau })\sim
[2\lambda_t\lambda_s|\hat{V}_{ub}|\cdot
|\hat{V}_{cb}|\epsilon ]^2~,
\label{width0}
\eeq
to be compared with the decay width of the dominant decay mode of minimal
SUSY $SU(5)$ \cite{his},

\beq
\Gamma (p\to K\nu_{\mu })_{SU(5)}\sim
[2\lambda_c\lambda_s\sin^2 \theta ]^2~,
\label{width1}
\eeq
where $\sin \theta $ is the Cabibbo angle.
For $\epsilon=0.22$, $|\hat{V}_{ub}|=0.0035$, $|\hat{V}_{cb}|=0.04$
(central values of CKM matrix elements \cite{dat}), from (\ref{width0}),
(\ref{width1}) we have

\beq
\frac{\Gamma (p\to K\nu_{\mu, \tau })}
{\Gamma (p\to K\nu_{\mu })_{SU(5)}}\simeq \frac{1}{300}~.
\label{rate}
\eeq
Therefore, in our model we expect
$\tau_p\sim 10^2\tau_p^{SU(5)}\sim 10^{32\pm 2}$~yr.
The decay modes into the charged leptons are more suppressed.


A crucial role in suppressing $d=5$ nucleon decay is played by the
mass scale $M_S$ of the symmetric $21$-plets, whose origin we do not explain.
In general, in $SU(5+N)$
GUTs even the origin of the GUT scale is unknown and only unification of
the three gauge couplings give information about its value. As we have seen,
the lower masses of additional symmetric plets can lead to greater nucleon stability.
The generation of mass scales is beyond the scope of
this paper.

Once we have insured that color triplet induced $d=5$ nucleon decay is
compatible with experiments, we should check whether the contribution from
Planck
scale induced operators are significant or not. In $SU(6)$
scenario $\frac{1}{M_P}qqql$-type operators arise from the coupling:

\beq
\frac{1}{M_P}\Gamma^{\alpha \beta \gamma \delta }
\frac{\Sigma^3 }{M_P^4}
21_{\alpha }\cdot 21_{\beta }\cdot 21_{\gamma }\cdot
\bar 6_{\delta }H~,
\label{plsc}
\eeq
where the $\Sigma$s appear due to symmetric $21$-plets, and  $\Gamma $
depends on powers of $X, H, \bar H$. For the operator $q_1q_1q_2l_2$,
which
gives the dominant contribution to nucleon decay,
$\Gamma^{1122}=\left(\frac{X}{M_P}\right)^7\left(\frac{\bar
HH}{M_P}\right)^2$. This, together with the suppression factors in
(\ref{plsc}), gives rise to a suppression
$\epsilon_G^3\epsilon^{12}$, which makes nucleon decay unobservable.
Another appropriate operator also can be obtained if instead of $21$s we use
$15$-plets in (\ref{plsc}) (this does not require the presence of $\Sigma$).
However, since $15_{1,2}$ contain $q_{1,2}$ states with weight
$\eta \sim 10^{-2}$, the suppression factor will be same.
As far as the `right'-handed operators
are concerned,
as was pointed out in \cite{str}, the dominant contribution arises
from the combination $u^ct^cd^c\tau^c $, either through chargino or wino
dressings. In our model operator responsible for coupling is

\beq
\left(\frac{X}{M_P}\right)^5\frac{\bar HH}{M_P^3}
15_1\cdot 20\cdot 20\cdot \bar 6_1H~.
\label{plscr}
\eeq
The diagrams which arise through chargino
dressings are irrelevant since there does not exist mixings of
$t^c$ state with the light generations
(see (\ref{up})). As far as the wino dressing diagrams
are concerned, the effective four-fermion $(u^cd^c)(s\nu_{\mu, \tau })$
operator
will have the coefficient $\epsilon^8\lambda_{\tau }V_{cb}$,
which provides strong suppression of nucleon decay.

        We therefore conclude that due to ${\cal U}(1)$ symmetry and
details of the $SU(6)$ model, Planck scale induced dimension five
operators are suppressed well beyond the required level.

\subsection{Neutrino Oscillations}

We now attempt to accommodate the recent solar and atmospheric neutrino
data (see \cite{sol} and \cite{atm} respectively) in the framework of the
PGB $SU(6)$ model. The prescription of ${\cal U}(1)$ charges
(see Table (\ref{t:fer})), permits us to
realize bi-maximal mixings between the active neutrinos,
and we will consider this
possibility in detail. Our strategy will be to follow the mechanism
suggested
in \cite{bimax}, in which such a scenario was realized in the framework
of $SU(5)$ which also gave a realistic pattern for charged fermion
masses and mixings.

The relevant couplings will be those which involve the
$(\bar 6+\bar 6')_{\alpha }$-plets. These contain the singlets
$(n+n')_{\alpha }$, and in order for the mechanism for maximal
neutrino mixings \cite{numssm} to work, they should decouple. For this,
we
introduce the states $(N+N')_{\alpha }$ with ${\cal U}(1)$ charges

$$
Q_{N_1}=Q_{N_1'}=2r'-r+3~,
$$
\beq
Q_{N_2}=Q_{N_2'}=Q_{N_3}=Q_{N_3'}=2r'-r+1~.
\label{Ncharges}
\eeq
Through the couplings

\beq
\bar 6_{\alpha }N_{\alpha }H+\bar 6_{\alpha }'N_{\alpha }'H~,
\label{6N}
\eeq
the states $(n+n')_{\alpha }$ form massive states with $(N+N')_{\alpha }$
respectively and will decouple at low energies.

Since we have assumed that the light $\nu_{\alpha }$
states reside in $\bar 6_{\alpha }$ (see section $3.2$), we will only
consider terms
involving them.
Because the ${\cal U}(1)$ charges of $\bar 6_2$ and $\bar 6_3$ are the same,
one can expect large $\nu_{\mu}-\nu_{\tau }$ mixings, which is desirable
for an explanation of the atmospheric anomaly. The ${\cal U}(1)$ charge of
$\bar 6_1$ state differs from those of $\bar 6_{2,3}$. This opens up
the possibility of obtaining maximal $\nu_e-\nu_{\mu, \tau }$
oscillations. To implement all of this, we introduce three right handed
states
${\cal N}_{1,2,3}$ with the following ${\cal U}(1)$
transformation properties:

\beq
Q_{{\cal N}_1}=-Q_{{\cal N}_2}=-r-2~,~~~~~
Q_{{\cal N}_3}=-r~,
\label{calcharges}
\eeq
and taking $r'=0$ in Table (\ref{t:fer}) , the relevant couplings
are

$$
\begin{array}{ccc}
 & {\begin{array}{ccc}
{\cal N}_1~~~~&\,\,{\cal N}_2~~~~&\,\,{\cal N}_3~~~~~~~~~~~~~~~~~
\end{array}}\\ \vspace{2mm}
\begin{array}{c}
\bar 6_1\\ \bar 6_2 \\ \bar 6_3

\end{array}\!\!\!\!\! &{\left(\begin{array}{ccc}
\,\, \left(\frac{X}{M_P}\right)^4~~ &
\,\, 1~~&
\,\, a\left(\frac{X}{M_P}\right)^2
\\
\,\, \left(\frac{X}{M_P}\right)^2~~ &
\,\,0~~&
\,\, b
\\
\,\, \left(\frac{X}{M_P}\right)^2~~ &
\,\,0~~&
\,\, c
\end{array}\right)\frac{XH}{M_P}(1+\frac{\Sigma }{M_P}) }~,
\end{array}  \!\!~~~
$$
\beq
\begin{array}{ccc}
 & {\begin{array}{ccc}
{\cal N}_1~~~~~&\,\,{\cal N}_2~~~~~&\,\, {\cal N}_3~~~~~~~
\end{array}}\\ \vspace{2mm}
\begin{array}{c}
{\cal N}_1 \\ {\cal N}_2 \\ {\cal N}_3

\end{array}\!\!\!\!\! &{\left(\begin{array}{ccc}
\,\, M'\left(\frac{X}{M_P}\right)^2 &
\,\, M' &
\,\, M''\left(\frac{X}{M_P}\right)^2
\\
\,\, M' &
\,\, 0 &\,\,0
\\
\,\, M''\left(\frac{X}{M_P}\right)^2 &
\,\, 0 & \,\, M
\end{array}\right)
\left(\frac{\bar HH}{M_P^2}\right)^2
}
\end{array}~~~
\label{Ns}
\end{equation}
where $a,b,c $ are dimensionless couplings of order unity. If for the
mass scales in (\ref{Ns}) we take $M'\sim M''\simeq 3\cdot 10^{16}$~GeV,
$M\simeq 10^{15}$~GeV, integration of ${\cal N}$ states leads to the
following mass matrix for the `light' neutrinos

\beq
\begin{array}{ccc}
 & {\begin{array}{ccc}
~& \,\,~  & \,\,~~
\end{array}}\\ \vspace{2mm}
\begin{array}{c}
\\  \\
 \end{array}\!\!\!\!\!\!\!\!&{\hat{m}_{\nu }=\left(\begin{array}{ccc}
\,\,a^2\epsilon^4  &\,\,~~ab\epsilon^2 &
\,\,~~ac\epsilon^2
\\
\,\,ab\epsilon^2   &\,\,~~b^2  &
\,\,~~bc
 \\
\,\, ac\epsilon^2 &\,\,~~bc  &\,\,~~c^2
\end{array}\right)m }~
\end{array}  \!\!  ~~+
\begin{array}{ccc}
 & {\begin{array}{ccc}
~~\,\,~  & \,\,~~~~
\end{array}}\\ \vspace{2mm}
\begin{array}{c}
\\  \\
 \end{array}\!\!\!\!\!\!\!\!&{\left(\begin{array}{ccc}
\,\,\epsilon^2 &\,\,~1 &
\,\,~1
\\
\,\,1   &\,\,~0  &
\,\,~0
 \\
\,\, 1 &\,\,~0  &\,\,~0
\end{array}\right)m' }
\end{array}  \!\!  ~
\label{mnu}
\end{equation}

where

\beq
m\equiv \frac{\epsilon^2 h_u^2}{M'}\simeq 4\cdot 10^{-5}~{\rm eV}~,~~~~
m'\equiv\frac{h_u^2}{M }\simeq 3\cdot 10^{-2}~{\rm eV}~.
\label{scales}
\eeq

From (\ref{mnu}), (\ref{scales}) one can verify that the `heaviest'  state
has mass

\beq
m_{\nu_3}\simeq (b^2+c^2)m\simeq 3\cdot 10^{-2}~{\rm eV}~,
\label{mnu3}
\eeq
while the two `light' states are nearly degenerate

\beq
m_{\nu_1}\simeq m_{\nu_2}\simeq m'\simeq 4\cdot 10^{-5}~{\rm eV}~.
\label{masdeg}
\eeq
Consequently, for the solar and atmospheric neutrino oscillation parameters
respectively, we find (taking into account that $b\sim c$):


$$
\Delta m^2_{21 }\sim 2m'^2\epsilon^2\simeq 10^{-10}~{\rm eV}^2~,
$$
\beq
\sin^2 2\theta_{e2,3} =1-{\cal O}(\epsilon^2)~,
\label{solosc}
\eeq
and

$$
\Delta m^2_{32}\simeq m_{\nu_3}^2\simeq m^2\sim 10^{-3}~{\rm eV}^2~,
$$
\beq
\sin^2 2\theta_{\mu 3}=\frac{4b^2c^2}{(b^2+c^2)^2}\sim 1~.
\label{atmosc}
\eeq
These suggest the solution of the solar neutrino puzzle through
maximal $\nu_e-\nu_{\mu, \tau }$ vacuum oscillations, while the
atmospheric neutrino anomaly is explained via large
$\nu_{\mu }-\nu_{\tau }$ mixings.

Although we have concentrated on the bi-maximal scenario, by proper
selection of the
right handed neutrino content and their mass scales, it is possible to
resolve the solar neutrino puzzle through the small angle MSW
oscillations. Also, it is possible to introduce a sterile neutrino state
(which can be kept light by the ${\cal U}(1)$ symmetry \cite{chun,
flipsu6, 422}) and
realize
different oscillation scenarios for a simultaneous explanation of the
solar and atmospheric neutrino data. For more details about all this we
refer the reader to
\cite{bimax}, where these issues are discussed.

\section{Conclusions}

In this paper we have proposed a rather general mechanism for suppressing
nucleon decay in SUSY GUTs which embed the MSSM gauge group in $SU(5+N)$.
This mechanism can be a powerful tool for realistic model
building. One particularly interesting transparent example is the
pseudo-Goldstone $SU(6)$ model, which we have investigated in detail.
We have observed that, by supplementing $SU(6)$ with an anomalous ${\cal
U}(1)$ flavor
symmetry, we can obtain a more stable nucleon, provide an
`all-order' resolution of the gauge hierarchy problem, a natural explanation
of the charged
fermion mass hierarchies and mixings, and obtain an acceptable value for the
strong coupling
$\alpha_s(M_Z)$. For solving the
solar and atmospheric neutrino puzzles, we have presented the `bi-maximal'
neutrino mixing scenario, which nicely blends with the pattern
of charged fermion masses and mixings.

\end{document}